%% file: pears.tex
\documentclass[twocolumn]{aastex631}

\usepackage{microtype}  
\usepackage{amsmath}
\usepackage{amsfonts}
\usepackage{amssymb}
\usepackage{multirow}
\usepackage{tikz}
\usepackage{xcolor}
\usepackage{soul}
\usepackage{hyperref}

\definecolor{pink}{RGB}{232,132,161}
\definecolor{yellow}{RGB}{255,213,0}

\newcommand{\ms}[1]{\ensuremath{M_{*{#1}}}}

\newcommand{\subcat}{\textit{Subhalo Catalog}}
\newcommand{\starcat}{\textit{Subhalo + Stellar Mass Catalog}}
\newcommand{\paircat}{\textit{Full Pair Catalog}}
\newcommand{\Rvir}{\ensuremath{\rm R_{vir}}}

\shorttitle{Pair fractions in TNG100}
\shortauthors{Chamberlain et al.}

\graphicspath{{./}{../plots/paper1/}}
\input{math_definitions.tex}

\newcommand{\affuofa}{University of Arizona, 933 N. Cherry Ave, Tucson, AZ 85721, USA}
\newcommand{\affunam}{Instituto de Radioastronom\'ia y Astrof\'isica, Universidad Nacional Aut\'onoma de M\'exico, Apdo. Postal 72-3, 58089 Morelia, Mexico}
\newcommand{\affuoff}{Department of Astronomy, University of Florida, P.O. Box 112055, Gainesville, FL 32611, USA}
\newcommand{\affberk}{Department of Astronomy, University of California, Berkeley, 501 Campbell Hall, Berkeley, CA 94720, USA}
\newcommand{\affuofu}{Department of Physics and Astronomy, University of Utah, 115 South 1400 East, Salt Lake City, UT 84112, USA}
\newcommand{\affnyu}{The Center for Cosmology and Particle Physics, New York University, 726 Broadway, New York, NY 10003}
\newcommand{\affnrao}{National Radio Astronomy Observatory, 520 Edgemont Road,
Charlottesville, VA 22903, USA}
\newcommand{\affuva}{Department of Astronomy, University of Virginia, 530 McCormick Road, Charlottesville, VA 22904, USA}

\newcommand{\affnottingham}{School of Physics and Astronomy, University of Nottingham, University Park, Nottingham NG7 2RD, UK}
\newcommand{\affkorea}{Korea Astronomy and Space Science Institute, 776 Daedeokdae-ro, Yuseong-gu, Daejeon 34055, Republic of Korea}
\newcommand{\afftrent}{Department of Physics and Astronomy, Trent University, 1600 West Bank Drive, Peterborough, ON K9L 0G2, Canada}
\newcommand{\affoccidental}{Occidental College, Physics Department, 1600 Campus Road, Los Angeles, CA 90041, USA}

\begin{document}

\title{
  A physically motivated framework to compare pair fractions\\ of isolated low and high mass galaxies across cosmic time
  }

\author[0000-0001-8765-8670]{Katie~Chamberlain}
\thanks{E-mail: katiechambe@arizona.edu}\affiliation{\affuofa}

\author[0000-0003-0715-2173]{Gurtina Besla}
\affiliation{\affuofa}

\author[0000-0002-9820-1219]{Ekta~Patel}
\thanks{Hubble Fellow}\affiliation{\affberk} \affiliation{\affuofu}

\author[0000-0002-9495-0079]{Vicente Rodriguez-Gomez}
\affiliation{\affunam}

\author[0000-0002-5653-0786]{Paul Torrey}
\affiliation{\affuoff}

\author[0000-0003-2939-8668]{Garreth Martin}
\affiliation{\affnottingham}\affiliation{\affkorea}\affiliation{\affuofa}

\author[0000-0001-8348-2671]{Kelsey Johnson}
\affiliation{\affuva}

\author[0000-0002-3204-1742]{Nitya Kallivayalil}
\affiliation{\affuva}

\author[0000-0002-1871-4154]{David Patton}
\affiliation{\afftrent}

\author[0000-0003-0256-5446]{Sarah Pearson}
\thanks{Hubble Fellow}\affiliation{\affnyu}

\author[0000-0003-3474-1125]{George Privon}
\affiliation{\affnrao}\affiliation{\affuoff}\affiliation{\affuva}

\author[0000-0002-2596-8531]{Sabrina Stierwalt}
\affiliation{\affoccidental}

\begin{abstract} 
    Low mass galaxy pair fractions are under-studied, and it is unclear whether low mass pair fractions evolve in the same way as more massive systems over cosmic time.
    In the era of JWST, Roman, and Rubin, selecting galaxy pairs in a self-consistent way will be critical to connect observed pair fractions to cosmological merger rates across all mass scales and redshifts.
    Utilizing the Illustris TNG100 simulation, we create a sample of physically associated low mass ($\rm 10^8<M_*<5\times10^9\,\Msun$) and high mass ($\rm 5\times10^9<M_*<10^{11}\,\Msun$) pairs between $z=0$ and $4.2$.
    The low mass pair fraction increases from $z=0$ to $2.5$, while the high mass pair fraction peaks at $z=0$ and is constant or slightly decreasing at $z>1$. 
    At $z=0$, the low mass major (1:4 mass ratio) pair fraction is 4$\times$ lower than high mass pairs, consistent with findings for cosmological merger rates. 
    We show that separation limits that vary with the mass and redshift of the system, such as scaling by the virial radius of the host halo ($r_{\mathrm{sep}}< 1 R_{\rm vir}$), are critical for recovering pair fraction differences between low mass and high mass systems.
    Alternatively, static physical separation limits applied equivalently to all galaxy pairs do not recover the differences between low and high mass pair fractions, even up to separations of $300\,\kpc$.
    Finally, we place isolated mass-analogs of Local Group galaxy pairs, i.e., Milky Way (MW)--M31, MW--LMC, LMC--SMC, in a cosmological context, showing that isolated analogs of LMC--SMC-mass pairs and low-separation ($<50\kpc$) MW--LMC-mass pairs are $2-3\times$ more common at $z\gtrsim2-3$.
\end{abstract}

\section{Introduction} \label{sec:intro}
    Galaxy mergers have been studied in detail as a mechanism for driving galaxy evolution, and have been identified as a trigger of, for example, 
    active galactic nuclei (AGN)~\citep[e.g.][]{Hopkins2008,Treister2010,Ramos2011,Satyapal2014,Comerford2015,Glikman2015,Blecha2018,Ellison2019}, 
    star formation~\citep[e.g.][]{Mihos1996, DiMatteo2008,Ellison2011,Hopkins2013,Patton2013,Martin2017,Hani2020,Patton2020,Martin2021}, 
    and changes in morphology~\citep[e.g.][]{Conselice2003,Lotz2008,Casteels2014,Patton2016,Bignone2017,RG2017,Martin2018,Jackson2019,Snyder2019,Jackson2022,Guzman-Ortega2023}. 
    Constraining the merger rate for galaxies is critical for quantifying the importance of mergers for galaxy evolution and testing predictions for hierarchical assembly in cold dark matter theory \citep[e.g.][]{Stewart2009,Hopkins2010,RG2015}.

    In practice, merger rates cannot be measured straightforwardly from observations and, rather, are calculated by converting the observed frequency of a galaxy merger signature (i.e., asymmetry, concentration, pair fraction, etc.) to a merger rate using an observability timescale~\citep[e.g.][]{Lotz2011}. However, morphological signatures of mergers are often caused by non-merger phenomena.
    For example, low mass galaxies are more commonly morphologically disturbed by fly-bys and non-merger interactions than by mergers~\citep{Martin2021}, 
    and star-forming galaxies at high redshift tend to be clumpy and asymmetric even in isolation~\citep{Wuyts2013}.
    Additionally, detection of these signatures can be strongly dependent on image depth or galaxy stellar mass, and identifying tidal features that rely on human classifications may be unreliable~\citep{Martin2022}.

    Pair fractions, on the other hand, can be calculated independently of morphological information, and thus offer a more robust method of observationally quantifying merger rates across cosmic time. 
    Pair fractions of high mass ($M_*\gtrsim10^{10}\Msun$) galaxies have been well studied across cosmic time both observationally~\citep[e.g.][]{Patton2002,Lin2004,Lin2008,Lotz2011,Ferreras2014, Man2016,Duncan2019} and theoretically~\citep[e.g.][]{RG2015,Snyder2017,Snyder2023}, with theoretical studies typically done in projection for comparison to observational campaigns\footnote{The studies cited here have limits on the projected separation of their sample, such that the projected separations range between 5 \kpc\  and $\sim140$ \kpc.}.  
    Low mass galaxy pairs ($10^8<M_*<5\times10^9\Msun$) have been studied at low redshift~\citep[e.g.][]{Stierwalt2015,Pearson2016,Besla2018,Paudel2018,Luber2022},
    but are less well understood across cosmic time owing to the difficulty in observing faint systems outside of the Local Volume. 
    However, JWST~\citep{Gardner2006}, as well as the next generation of deep and wide-field surveys from Rubin Observatory\footnote{ \href{https://rubinobservatory.org/}{https://rubinobservatory.org}}~\citep{RUBIN2019} and Roman Space Telescope\footnote{\href{https://roman.gsfc.nasa.gov/}{https://roman.gsfc.nasa.gov}}~\citep{Spergel2015}, will significantly revolutionize our ability to identify such systems at high redshift $(z\gtrsim10)$~\citep{Behroozi2020} and in abundance at lower redshift ($z\lesssim6$)~\citep{Robertson2019a,Robertson2019b-LSST}.
    
    There are reasons to believe that low mass and high mass pairs evolve differently as a function of time.
    For example, semi-empirical and cosmological studies find that galaxy merger rates vary both with redshift and the mass of the most massive galaxy of the pair~\citep[see, e.g.][]{Guo2008,Stewart2009,Hopkins2010,RG2015,Martin2021}.
    It is thus reasonable to assume that the pair fractions of these two mass scales reflect these evolutionary differences as well. 

    The redshift evolution of low and high mass pair fractions has not yet been studied in simulations in a self-consistent way, where high mass and low mass pairs are selected from simulations using otherwise equivalent selection criteria. 
    We aim to characterize the redshift behavior of pair fractions of low mass and high mass pairs, independent of environmental and projection effects, and to create a robust framework in which to fairly compare pair fractions of different mass scales across cosmic time. 

    In particular, we take the approach of consistently selecting physically associated pairs of low mass ($10^8<M_*<5\times10^9\Msun$) and high mass $(5\times10^9<M_*<10^{11}\,\Msun)$ galaxies. 
    Specifically, we identify major ($1-1/4$) and minor $(1/10-1/4)$ stellar mass ratio pairs from $z=0$ to $z=4.2$ in the IllustrisTNG cosmological simulation, TNG100. 
    We require that pairs are part of the same Friends of Friends (FoF) group and that no other more massive perturbers are nearby, which allows us to ensure that the recovered pair fractions are inherent to the population of selected pairs rather than a function of environment. 

    Typically, pair fraction studies via simulations and observations apply physical projected separation cuts that are constant over time and do not vary with the mass of the target system. 
    In this study, we identify how the application of a fixed separation criterion affects inferred pair fractions, and show that a time and mass evolving separation cut is necessary to permit equitable pair fraction comparisons across different mass regimes. 
    Although implementing alternative selection criteria in future observational pair fraction studies may not always be strictly necessary, we recommend that pair selection criteria be reevaluated for observational studies that seek to compare pair fractions as a function of mass or redshift.

    Finally, there are a number of galaxy pairs in the Local Group that are mass analogs of the isolated pairs in our sample.
    For example: the MW--M31 system is a high mass major pair, the MW--LMC and M31--M33 systems are high mass minor pairs, and the LMC--SMC are a low mass minor pair.
    Studies have examined the frequency of such configurations, particularly at low redshift or in the form of progenitor systems of the Local Group in cosmological simulations~\citep{Bk2011,Fattahi2013,Patel2017a-Orbits,Geha2017,Mao2021}, but the prevalence of such pairs in isolation has not yet been quantified as a function of redshift. 
    We utilize our pair dataset to quantify the likelihood of finding isolated mass-analogs of Local Group pairs as a function of redshift, particularly when the present day separations of these systems are folded in. 
    
    This paper is structured as follows. 
    In Section~\ref{sec:methods}, we outline our methodology for selecting physically associated high mass and low mass pairs from the TNG simulation. 
    In Section~\ref{sec:pairprops}, we provide an overview of the properties of the selected sample, including the number of primaries, pairs, and their stellar mass ratios as a function of time. 
    We present the time-evolving pair fraction of high mass and low mass pairs in Section~\ref{sec:results}, and show how they change for different separation criteria. 
    In Section~\ref{sec:discussion}, we give context to our results by drawing comparisons to Local Group pairs and other Illustris-based pair fraction studies, and discuss implications for observational campaigns.
    Finally, we summarize our results and conclusions in Section~\ref{sec:summary}.

 
\section{Methodology}\label{sec:methods}
We aim to quantify and characterize the frequency, or pair fraction, of low mass ($\rm\ms{}=10^{8}-5\times 10^{9}\Msun$) and high mass ($\rm\ms{}=5\times10^{9}-10^{11}\Msun$) galaxy pairs in cosmological simulations as a function of cosmic time. 
To this end, we utilize the \tng\ suite of simulations to select subhalo pairs as a function of redshift, according to the selection criteria outlined in the remainder of this section. 

In Sec.~\ref{sec:methods-sims}, we provide motivation for and details of the simulation utilized in this study.
Sec.~\ref{sec:methods-halos} outlines the initial mass cuts used to define our \subcat, to which we will add stellar mass information. 
Sec.~\ref{sec:methods-am} outlines the abundance matching prescription used to associate dark matter subhalos with stellar masses, and the creation of the \starcat, from which we will select pairs.
Sec.~\ref{sec:methods-pairs} describes the second set of selection criteria that we use to finally construct the \paircat. 

    \subsection{Simulation details} \label{sec:methods-sims}
    The IllustrisTNG project~\citep{TNG1, TNG2, TNG3, TNG4, TNG5} consists of a suite of dark-matter-only $N$-body and full physics cosmological simulations that adopt the \textit{Planck2015} \lcdm\  cosmology~\citep{Planck2015}.
    
    In this study, we utilize data from TNG100-1, the main high-resolution, full physics simulation of a $(110.7\,\Mpc)^3$ volume (hereafter TNG100)\footnote{We will also use data from TNG100-1-Dark, the main high-resolution dark-matter-only run (hereafter TNG100-Dark), as presented in Sec.~\ref{sec:disc-dark}.} 
    This simulation follows the evolution of baryons and $1820^3$ dark matter particles from $z=127$ to $z=0$.  
    The volume of this simulation is sufficiently large, and the resolution is sufficiently high, to conduct a simultaneous analysis of the statistics of both low mass and high mass galaxy pairs, as shown by studies of pair statistics in similarly sized volumes~\citep{Sales2013,Patel2017a-Orbits,Patel2017b-Masses,Besla2018}. 

    We utilize the group catalogs produced by the \texttt{SUBFIND} algorithm~\citep{Springel2001,Dolag2009}. 
    This catalog was created using the Friends-of-Friends (FoF) algorithm~\citep{Davis1985}, which links nearby dark matter particles to define large halos of associated particles, from which subhalos are identified as overdense and gravitationally bound dark matter structures.
    In addition, we utilize the merger trees provided by the \texttt{SUBLINK} algorithm~\citep{RG2015}, which tracks subhalos between snapshots, enabling us to trace the mass evolution of our selected subhalos. 


    \subsection{Choosing low and high mass subhalos in TNG100} \label{sec:methods-halos}

    In this section, we outline the steps to create the \subcat, which we will utilize to assign stellar masses to each of the dark matter subhalos. 

    To assess whether the pair fractions of low mass and high mass pairs evolve in fundamentally different ways as a function of redshift, rather than as a product of their environmental conditions, we will focus on galaxy pairs in low-density environments only.\footnote{Note that our pair fraction calculations are thus specifically for isolated systems, and the global pair fraction (including both isolated and non-isolated pairs) will be different.}
    As such, we ensure that our target subhalos are sufficiently isolated by placing limits on the virial mass of the FoF group to which our low mass or high mass samples belong.    

    The following selection process is repeated for each snapshot of the simulation over a redshift range of $z=0-4.2$.  
    We stop our analysis at $z=4.2$ since the population of massive subhalo pairs falls off rapidly at larger redshifts, where the sample size of high mass primaries falls well below 100 subhalos per snapshot.
    
    At each snapshot, we first apply a cut on the FoF group virial mass, $M_{\rm vir}$, given by \texttt{Group\_M\_TopHat200} in the TNG100 group catalogs.\footnote{The mass \texttt{Group\_M\_TopHat200} is the mass enclosed by a sphere with mean density $\Delta_c *\rho_c$, where $\Delta_c$ is the overdensity constant from~\citet{Brynorman1998} and $\rho_c$ is the critical density of the universe at the time calculated. The corresponding virial radius in TNG100 is given by \texttt{Group\_R\_TopHat200}.}
    We define the group mass range for low mass and high mass groups as: 
    \begin{align*}
        \mbox{\textbf{low mass:}}&\,\rm M_{G} = 8\times 10^{10}- 5\times 10^{11}\,\Msun \\ 
        \mbox{\textbf{high mass:}}&\, \rm M_{G}=10^{12}- 6.5\times10^{12}\,\Msun.
    \end{align*}
    The FoF group mass criteria is fixed for all redshifts, which means that some low mass groups at high $z$ may be the progenitors of high mass systems at $z=0$.

    By requiring that the FoF group virial mass does not exceed the above limits, we ensure that there are no subhalos more massive than these limits that will perturb the dynamical state of identified pairs. 
    For example, selecting low mass pairs from the low mass FoF groups ensures that the selected pairs are not satellite systems of high mass subhalos. 
    Over 99\% of subhalos selected from FoF groups in these mass limits are not within $1.5\,\Mpc$ of a more massive perturber at $z=0$.

    From the set of FoF groups that pass the group mass cut, we create a catalog of all subhalos within each FoF group that pass a current ``minimum mass" threshold.  
    At the given snapshot, we require a minimum subhalo mass, \Mhalo, to be:
    \begin{equation*}
    \mbox{\textbf{minimum subhalo mass:}}\,
    \Mhalo > 1\times10^{9}\Msun.
    \end{equation*}
    \Mhalo\ is given by the \texttt{SubhaloMass} field in the group catalogs, and is the total combined mass of all bound dark \textit{and} baryonic matter.
    This ``minimum mass" threshold ensures that subhalos are resolved into $>100$ dark matter particles, and thus should be robustly identifiable using the \subfind\ and \sublink\ algorithms~\citep{RG2015}. 

    For each subhalo that passes both criteria, i.e. the subhalo belongs to either a high mass or low mass FoF group and also passes the ``minimum mass" threshold, we use the \sublink\ merger trees to identify the maximum subhalo mass achieved by the subhalo, \Mpeak. 
    We consider only the given snapshot and all previous snapshots in the determination of \Mpeak.

    All subhalos in a given snapshot that pass both the FoF group and ``minimum mass" selection criteria form our full sample of subhalos, which we call the \subcat. 
    The catalog contains subhalos at each snapshot from $z=0$ to 4.2 (snapshot numbers 20-99) and their associated properties, namely: Subhalo ID, FoF Group Number, current subhalo mass \Mhalo, and peak subhalo mass \Mpeak. 
    At z=0, this selection process results in 44,656 subhalos in low mass groups and 38,350 subhalos in high mass groups.   
    
    From here, we will use additional selection criteria to identify pairs of subhalos in each group.

    \subsection{Abundance matching} \label{sec:methods-am}
    We utilize a stellar mass to halo mass (SMHM) relationship to assign stellar masses to each of the subhalos in our \subcat.  
    There are a few reasons we opt to assign stellar masses to our subhalos, rather than use those computed directly from the stellar particles in TNG100.

    First, and primarily, utilizing an abundance matching prescription enables the direct comparison of results between dark-matter-only and full hydrodynamics simulations, since the stellar masses are assigned in an identical and prescriptive way. 
    We make a brief note of results for our equivalent analysis using the dark-matter-only, TNG100-Dark simulation, in Sec.~\ref{sec:disc-dark}.
    
    Second, while the SMHM function of \tng{} at $z=0$ closely reproduces the profile of the SMHM relation from various abundance matching and semi-empirical models~\citep{TNG5,Nelson2019}, using stellar masses as calculated from abundance matching allows us to avoid any simulation-dependent stellar mass effects.
    In particular, we would like to avoid a dependence between our results and the particular subgrid physics model implemented in \tng{}.
    
    Third, abundance matching allows us to account for the observed spread in the SMHM relationship, as we can sample the relation many times to get a distribution of stellar masses for each subhalo in the simulation. Otherwise, we would only be able to perform this analysis once given the set of stellar masses from the simulation. 
    This is particularly important in the low mass regime ($M_h \lesssim 10^{11}\Msun$) where the scatter in the SMHM function is large, between $\sim 0.3$ for $M_{h}\sim 10^{10}\Msun$ up to $\sim 1$dex for $M_{h}\sim10^{8}\Msun$~\citep{Munshi2021}.
    
    We use the abundance matching relationship presented in~\citet{Moster2013}.
    The SMHM relationship therein provides an analytic prescription to assign stellar masses to dark matter halos as a function of subhalo mass and redshift, and includes terms to account for the systematic scatter in the SMHM relationship, with a larger scatter at lower halo masses.
    We were careful to choose the input subhalo mass and redshift that would assign the most accurate stellar masses to each halo given their individual histories.
    
    The abundance matching prescription is calibrated for centrals of FoF groups; thus, we elect to use the maximum subhalo mass \Mpeak\ in the stellar mass calculation, rather than the subhalo mass at the given snapshot \citep[see][]{Besla2018}.
    Using \Mpeak\ mass allows us to remain robust to scenarios in which a secondary has formed most of its stars, then loses a significant portion of its dark matter content through tidal interactions with a primary, but retains the bulk of its stellar content.
    In fact, \citet{Munshi2021} found that the stellar mass of subhalos at $z=0$ in the ``Marvel-ous Dwarfs" and ``DC Justice League" zoom simulations are more closely correlated with $M_{\mathrm{peak}}$ than the $z=0$ halo mass for halos with peak halo mass $10^8<M_{\rm peak}<10^{11} \Msun$. 

    We also use the current redshift of the given snapshot, $z_{\mathrm{snap}}$, of each subhalo in the stellar mass calculation. 
    Using $z_{\mathrm{snap}}$ means that we account for changes in the stellar mass of both the primary and secondary halo, even after the secondary has entered the primary's halo. 
    This assumption is consistent with findings from~\cite{Akins2021}, which found that massive dwarf satellites ($M_*\approx 10^8-10^9\, \Msun$) entering MW-mass host halos are rarely quenched, and with~\cite{Geha2013}, which found that dwarfs $>1.5\,\Mpc$ from a MW-type galaxy are often star forming and rarely quenched.
    Additionally, the SAGA survey has found that large satellites of MW-type galaxies are often very blue, with infall into the halo spurring high rates of star formation due to the large gas fraction in dwarfs~\citep{Mao2021}. 
    Thus, our abundance matching prescription is of the form $\ms{}=f(\Mpeak,z_{\mathrm{snap}})$. 

    We calculate the stellar mass for a given subhalo by Gaussian sampling each of the fitting parameters of the analytic framework from~\cite{Moster2013}, in order to account for the spread in the SMHM relationship.
    We generate a single stellar mass ``realization" using an independent draw from the SMHM distribution to calculate a stellar mass for each subhalo of the full \subcat.
    For each snapshot, we repeat this process 1000 times to generate 1000 separate realizations of assigned stellar masses for the \subcat.
    The resulting catalog is called the \starcat, and consists of the set of all subhalos from the \subcat, as well as the 1000 stellar mass realizations. 
    Each realization is treated as an independent sample of galaxy stellar masses, which will allow us to report realistic spreads of pair properties. 


    \subsection{Pair selection}\label{sec:methods-pairs}
    Starting from the \starcat, we outline the pair matching process used to generate the \paircat below. 
    At each redshift, and for each stellar mass realization, we will identify subhalo pairs consisting of a ``primary" and a ``secondary," where primaries are the more massive of the pair by stellar mass.

    \subsubsection{Selecting primaries}
        Primary galaxies (equivalently, ``primary subhalos") are the most massive galaxy of their FoF group by stellar mass, such that each group will have a singular primary galaxy.  
        Each stellar mass realization is treated independently, and so the subhalo identified as the primary of a group may change between stellar mass realizations. 

        At a given snapshot, and for each stellar mass realization, we rank-order the subhalos of each FoF group by stellar mass. 
        Primaries are defined as the subhalo with the highest stellar mass ($M_{*1}$) in their FoF group that passes the following criteria:         
        \begin{align*} 
        \mbox{\textbf{low mass primaries:}}&\, 10^{8}< \rm M_{*1} < 5\times10^{9} \Msun \\ 
        \mbox{\textbf{high mass primaries:}}&\, 5\times 10^{9}< \rm M_{*1} < 10^{11} \Msun.
        \end{align*}
        The stellar mass criteria is fixed, and \textit{does not change} as a function of redshift. 
        At $z=0$, the stellar mass range for low mass primaries corresponds to isolated analogs of the LMC or M33, while the high mass primaries represent isolated analogs of the MW or M31. 

        Since our selection is based on stellar mass, and there is a large spread in the SMHM relation, a primary subhalo may not be the most massive subhalo in terms of total subhalo mass. 

    \subsubsection{Selecting secondary companions}
        As before, the selection of secondary companions occurs independently for each stellar mass realization, and for each snapshot. 
        Secondary subhalos are defined as the second most massive subhalo in a FoF group by stellar mass ($M_{*2}$). 
        Secondary subhalos must also have a stellar mass ratio with respect to primaries of:
        \begin{align*}
            &\mbox{\textbf{stellar mass ratio:}}\,      
            M_{*2}/M_{*1} > 1/10.
        \end{align*} 

        We do not include companions with a stellar mass ratio $M_{*2}/M_{*1} <1/10$ as we will be limiting our pair sample to traditional definitions of major and minor pairs, 
        which are typically defined to have stellar mass ratios $M_{*2}/M_{*1} > 1/10$~\citep[i.e.][]{Lotz2011, RG2015, Snyder2017, Duncan2019, Wang2020, Guzman-Ortega2023}.

        Our subhalo ``minimum mass" threshold of $\Mhalo=10^9\, \Msun$ (described in Sec.~\ref{sec:methods-halos}) corresponds to a mean stellar mass of approximately $10^6\,\Msun$ at $z=0$. 
        This stellar mass is well below the 1/10 criteria for the lowest stellar mass primary considered ($M_{*1} > 10^8 \, \Msun$), ensuring that our pair sample will be complete even at the lowest stellar masses considered. 

    \subsubsection{Creating the pair catalog}
        Pairs consist of a single primary and secondary in a FoF group as defined via the criteria above, such that only one pair is identified in a single FoF group. 
        Each pair is categorized as either a major or minor pair based on the stellar mass ratio between the primary and secondary. Following, e.g. \cite{Lotz2011, RG2015}, 
        the stellar mass ratio criteria for major and minor pairs are:
        \begin{align*} 
        \mbox{\textbf{major pair:}}&\, \frac{1}{4}\leq \frac{\rm M_{*2}}{\rm M_{*1}}< 1 \\ 
        \mbox{\textbf{minor pair:}}&\, \frac{1}{10}\leq \frac{\rm M_{*2}}{\rm M_{*1}} < \frac{1}{4}.
        \end{align*}

        For each pair, we calculate the physical separation (proper \kpc, not comoving \kpc) between the two subhalos using the \texttt{SubhaloPos} field from the subhalo catalogs. We require that each pair have a minimum separation of $10 \, \kpc$ between primary and secondary to limit the impact of very close subhalos becoming indistinguishable in \subfind\ as a result of the resolution limit due to the softening length.

        If a primary subhalo does not have a companion that meets the stellar mass ratio criteria and separation criteria, the subhalo will still be considered a primary. 
        We refer to all primaries that do not have a companion with these criteria as ``isolated primaries," including primaries with a companion that meets the stellar mass ratio criteria but has a separation $<10\,\kpc$.
        The total number of primaries includes both isolated primaries and those with selected companions and is larger than the total number of pairs.

        For each snapshot and each of the 1000 realizations of the stellar mass in \starcat, we identify a set of isolated primaries and pairs. 
        Additionally, within each realization, no single subhalo can be a part of two separate pairs, such that the primary and secondary of every pair are unique to the pair.
        The final \paircat\ is the collection of all isolated primaries and pairs at each snapshot and includes the following information. 
        For each isolated primary, we store the current subhalo mass (\Mhalo) and stellar mass from the given realization. 
        For each pair, we store the primary and secondary subhalo masses (\Mhalo) and stellar masses, the pair separation, and the virial radius of the FoF group (see Sec.~\ref{sec:methods} for virial definitions).
        
\begin{figure*}[htb]
    \centering
    \includegraphics[width=\textwidth]{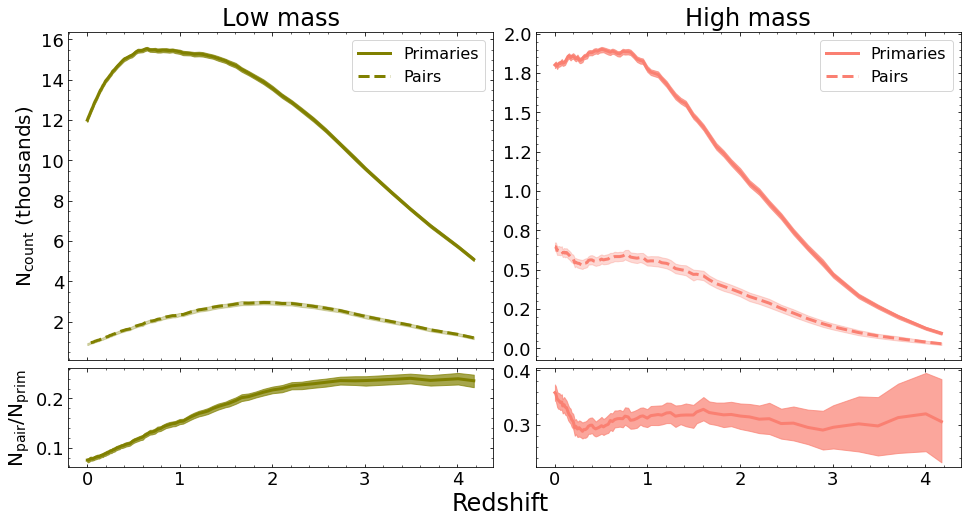}
    \caption{(Top) Number of isolated low mass (left) and high mass (right) primaries and pairs as a function of redshift.
    The solid and dashed lines represent the median of the set of total counts from each of the 1000 stellar mass realizations in the \paircat, while the shaded regions depict the 1st-99th percentile spread of the median.
    There are approximately 8 times as many low mass primaries as high mass primaries. 
    The low mass primary count (left) peaks at $z=1$ (with $\sim 15,000$ primary subhalos per realization), while the low mass pair count peaks at $z=2$ (with $\sim 3000$ pairs per realization). 
    The high mass primary count (right) peaks at $z\sim1$ (with $\sim 1900$ primary subhalos per realization), while the high mass pair count peaks at $z=0$ (with $\sim 700$ pairs per realization). 
    (Bottom) Total pair fraction (fraction of primaries with a major or minor secondary) as a function of redshift (see Sec.~\ref{sec:pairprops} for calculation details).
    The low mass total pair fraction is approximately flat between $z=2.5-4$, and decreases from $z=2.5$ to $z=0$. 
    The high mass total pair fraction is flat or decreasing from $z=1$ to $z=4$, but peaks sharply between $z=0-0.25$.}
    \label{fig:counts}
\end{figure*}

\begin{figure*}[htb]
    \centering
    \includegraphics[width=\textwidth]{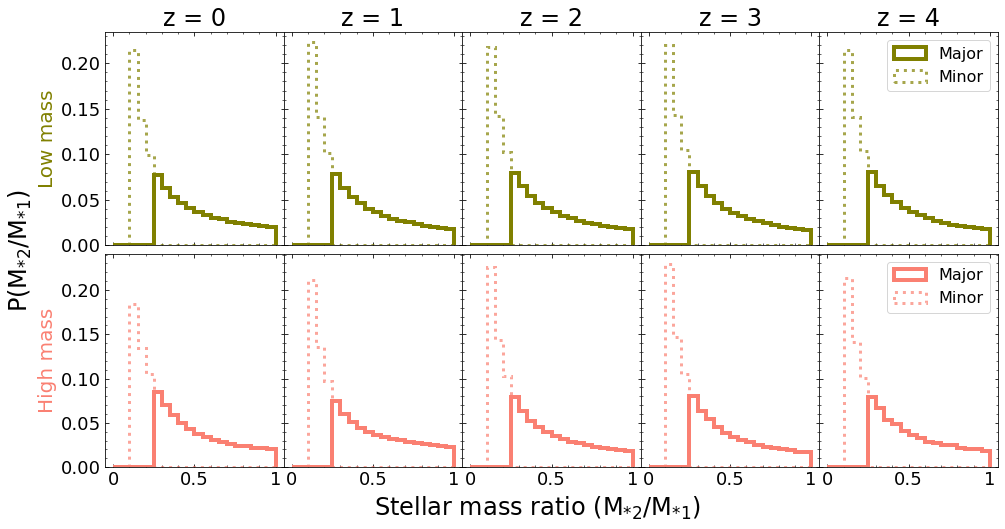}
    \caption{Stellar mass ratio distribution of all low mass (top) and high mass (bottom) pairs for all 1000 stellar mass realizations combined at each redshift. Major pairs (solid lines) are defined as pairs with mass ratio $\ms{2}/\ms{1} > 1/4$, while minor pairs (dotted lines) are defined as pairs with stellar mass ratio $1/10<\ms{2}/\ms{1}<1/4$. Overall, the stellar mass ratio distribution of major and minor pairs of low and high mass galaxies show little evolution from $z=4$ (right) to $z=0$ (left). 
    Major pairs make up $51-55\%$ of the total sample of pairs at every redshift for both low and high mass pairs.}
    \label{fig:massratio}
\end{figure*}

\section{Sample: Overview of pair properties} \label{sec:pairprops}
    Utilizing the \paircat, for each snapshot we compute the total number of primaries (isolated and paired) and pairs (including major and minor) in each of the 1000 realizations. We then compute the median and 1st and 99th percentile spread on the median over all realizations.\footnote{The spread on the median of each realization is very small. Thus, we opt to show the 1st and 99th percentile spread rather than those which align with traditional definitions of 1$\sigma$ or 2$\sigma$ measurements}
    Additionally, we compute the low and high mass total pair fraction for each individual realization, defined here as the ratio of the total number of pairs ($N_{\mathrm{pairs}}$) to the total number of isolated \textit{and} paired primaries ($N_{\mathrm{primaries}}$):
    $$f_{p}=\rm \frac{N_{\mathrm{pairs}}}{N_{\mathrm{primaries}}}.$$
    We again compute the median and spread over all 1000 realizations.
    
    Fig.~\ref{fig:counts} shows the median number (solid and dashed lines) of identified low and high mass primaries and pairs over the redshift range $z=0-4$.   
    The shaded regions show the 1-99\% spread of the set. 

    The number of identified primaries is lowest at $z=4$, and rises to a maximum around $z=1$ for both low and high mass primaries.
    The median count of low mass primaries (green solid line in top left panel) reaches a maximum of $15,545^{+89}_{-88}$ halos at $z\sim0.6$, then decreases by $\sim22\%$ to $12,002^{+85}_{-86}$ halos at $z=0$. 
    The count of high mass primaries (pink solid line in top right panel) reaches a maximum of $1901^{+16}_{-19}$ at $z=0.5$, and slightly declines to $z=0$. 
    Our sample of high mass primaries represents approximately 20\% of all subhalos in TNG100 with the same range of stellar masses at $z=0$.
    There are roughly 8 times as many low mass primaries as high mass primaries. Note that the comoving volume of \tng{} is the same at all redshifts.

    Unlike the primary count, the pair counts for low and high mass pairs peak at very different redshifts. 
    The count of low mass pairs (green dashed line) peaks much earlier, at $z=1.9$ with $2956^{+99}_{-95}$ pairs, and decreases to $896^{+51}_{-50}$ pairs at $z=0$.
    The pair count for high mass galaxies (pink dashed line), on the other hand, behaves more similarly to the primary count, increasing from $z=4$ to $z\sim1$, then decreasing to $z=0.25$ before peaking with $647\pm27$ pairs at $z=0$. 

    The bottom panel of Fig.~\ref{fig:counts} shows the total pair fraction for low mass and high mass pairs, or equivalently, the fraction of primaries with a major or minor companion.
    The total pair fractions for both low and high mass pairs are roughly flat for $z=2.5-4$, and display opposite behavior for low redshifts between $z=0-2.5$. 
    The low mass total pair fraction decreases from $0.233^{+0.008}_{-0.009}$ to $0.075\pm 0.004$, a decrease of roughly $68\%$, while the high mass total pair fraction remains flat between $z=1-2.5$, ranging between $0.275$ and $0.351$.
    At very low redshifts, from $z=0-0.25$, the high mass total pair fraction spikes sharply from $0.288^{+0.016}_{-0.013}$ to $0.359^{+0.015}_{-0.014}$, an increase of $37\%$.

    In Fig.~\ref{fig:massratio}, we show the combined distribution of stellar mass ratios of every pair from all 1000 realizations in the \paircat.
    Major pairs make up $51-55\%$ of the full sample of pairs at every redshift for both low and high mass pairs.
    In general, the shape of the distribution remains constant from $z=0$ (left) to $z=4$ (right) for both low and high mass pairs, and changes weakly as a function of mass scale and of redshift. 
    There are between $\sim3.3$ and $4.8$ times more pairs with mass ratios $\sim1/4$ than $\sim1/1$, and about $1.8-2.2$ times more pairs have $1/10$ than $1/4$. 
    Roughly 2\% of the total pair population is a 1:1 mass ratio encounter; this is true for low and high mass galaxies and across all redshifts considered.  

\section{Results: The frequency of low mass and high mass pairs}
\label{sec:results}

We have created catalogs of isolated low and high mass galaxy pairs from $z=0$ to 4.2 in the TNG100 simulation. 
In this section, we will analyze the frequency of major and minor pair types across cosmic time, with the goal of identifying potential differences between high and low mass galaxies. 
In Sec.~\ref{sec:results-frac}, we examine the redshift evolution of the fraction of primaries with a major or minor companion (the ``pair fraction") and compare the results for low mass and high mass pairs.
In Sec.~\ref{sec:results-frac-cuts}, we examine the redshift evolution of the pair fraction as a function of pair separation. 
In Sec.~\ref{sec:disc-dark}, we briefly describe how our equivalent pair fraction analysis utilizing the TNG100-Dark simulation compares to the TNG100 results presented in the previous two sections. 

In the following analysis, we will treat each stellar mass realization in the \paircat\ as an independent sample. 
At each snapshot, we calculate the median and 1st and 99th percentile spread on the median of the pair fractions in each of the 1000 realizations of the target pair sample.
Each of the following figures shows the median as the solid or dashed lines, and the shaded regions correspond to the spread on the median. 

    \subsection{Major and minor pair fractions}\label{sec:results-frac}
    The major and minor pair fractions are computed in the way as defined in Sec.~\ref{sec:pairprops} Eq.~\ref{eq}, where $N_{\rm pairs}$ is the total number of major \textit{or} minor pairs in the sample.
    For example, the low mass major pair fraction is the number of low mass major pairs divided by the number of low mass primaries, which can also be interpreted as the likelihood of finding a major companion of an isolated low mass primary. 
    
\begin{figure*}[htp]
      \centering
      \includegraphics[width=\textwidth]{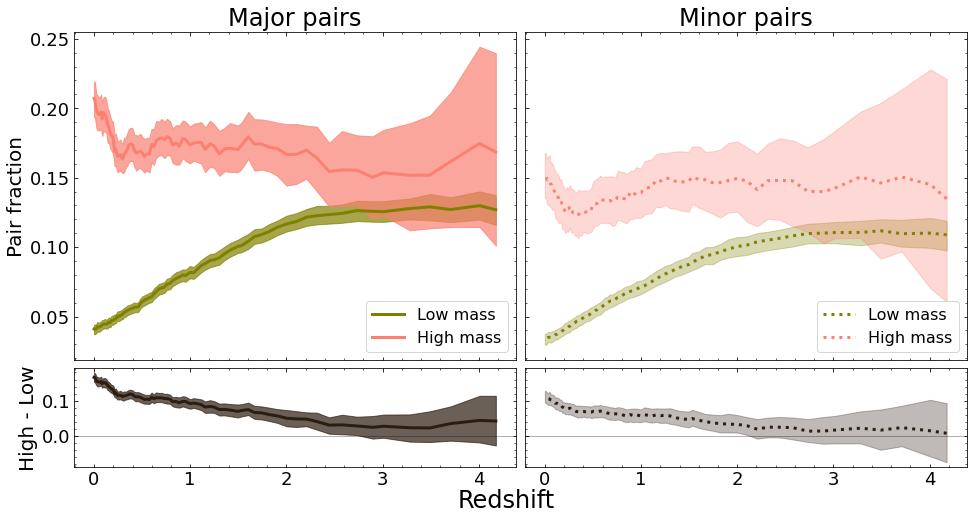}
      \caption{
        (Top) Median pair fraction as a function of redshift, defined as the fraction of low mass or high mass primaries with a major (solid) or minor (dashed) companion  (see Sec.~\ref{sec:methods-pairs}). All pairs in the \paircat\ must have a minimum separation of at least $>10\kpc$, with no constraint on the maximum separation.
        Shaded areas show the 1st-99th percentile range on the median (solid and dashed lines) from 1000 stellar mass realizations, as discussed at the beginning of Sec.~\ref{sec:results}. 
        Low mass major and minor pair fractions (green) are both at their minima at $z=0$, and increase by about 200\% by $z\sim2-2.5$, at which point they level off and remain constant from $z=3$ to 4. 
        On the other hand, high mass major and minor pair fractions reach their maxima at $z=0$, then abruptly decline until $z\sim0.25$ before remaining approximately constant from $z=1-4$. 
        (Bottom) The median and $1$st-$99$th percentile range of the subtracted difference between high and low mass pair fractions. 
        The difference peaks at $z=0$ for both major and minor pairs, and declines with increasing redshift.
        This panel shows that the redshift evolution of the pair fractions of low and high mass pairs proceeds differently, particularly at low redshift where pairs are more common for high mass galaxies than low mass galaxies.}
      \label{fig:pairfrac}
    \end{figure*}

    Fig.~\ref{fig:pairfrac} shows the pair fractions calculated for low and high mass pairs (including both major and minor pairs separately) as a function of time for $z=0-4$. 
    Note that, as discussed in Sec.~\ref{sec:methods-pairs}, we adopt a lower stellar mass ratio floor of $1/10$, and consequently the total pair sample is dominated by major pairs. 
    
    The low mass pair fraction evolves distinctly from that of high mass pairs.
    Low mass pair fractions for both major and minor pairs (green) are approximately constant from $z=3$ to 4, then decline monotonically to $z=0$. At $z=0$, the major pair fraction is $0.041^{+0.003}_{-0.004}$, while the minor pair fraction is $0.034\pm0.004$.
    At $z=3$, the pair fractions have increased by $207\%$ to $0.126^{+0.008}_{-0.007}$ for major pairs and $0.111\pm0.008$ for minor pairs. 

    High mass pair fractions (pink) remain approximately constant for $z>1$, where the median of the major pair fraction fluctuates between 0.150-0.179, and the median for minor pairs fluctuates between 0.135-0.151.
    Between $z=0.3$ and $z=0$, the major pair fraction increases from $0.163^{+0.011}_{-0.010}$ to $0.207^{+0.012}_{-0.014}$, while the minor pair fraction increases from $0.125^{+0.017}_{-0.014}$ to $0.152^{+0.016}_{-0.017}$.
    
    The bottom panels of Fig.~\ref{fig:pairfrac} show the low mass pair fraction subtracted from the high mass pair fraction (labeled ``$\rm High-Low$").
    The difference between high and low mass pair fractions increases with decreasing redshift, peaking at $z=0$ with a difference of $0.166\pm0.013$ for major pairs and $0.111\pm0.017$ for minor pairs. 
    Thus, at low redshift we expect both major and minor pairs to be more common in high mass galaxies than in low mass galaxies. 
    From $z=2.5$ to 4, the difference is approximately 0, and thus both major and minor pairs of high and low mass galaxies are equivalently common at high redshift. 
    
    Overall, these results show that low mass and high mass pair counts evolve differently over time, particularly at very low redshift, despite the pair fractions being roughly equal at higher redshift. 
    The implications for the difference in the evolution of pair fractions for low mass and high mass pairs across time are discussed in detail in Sec.~\ref{sec:discussion}.
    
\subsection{Major pair fractions as a function of separation}\label{sec:results-frac-cuts}
    In this section, we analyze subsets of the low and high mass major pairs from the previous section that pass additional separation criteria. 
    We study two different sets of separation criteria to compare the resulting pair fractions to the full sample shown in Fig.~\ref{fig:pairfrac}.
    In  Sec.~\ref{sec:results-frac-vircut} we apply a separation criterion that requires secondaries to be within a given factor of the virial radius of the FoF group, which is a reasonable proxy for the virial radius of the primary. 
    In Sec.~\ref{sec:results-frac-sepcut}, we use a range of limits on the 3D pair separation, where limits are consistent with values adopted in the literature. 
    
    In all cases, solid lines correspond to the median pair fraction and shaded regions to the 1st--99th percentile, as explained at the beginning of Sec.~\ref{sec:results}. 

\subsubsection{Pair separation limits as a function of the virial radius of the FoF group}\label{sec:results-frac-vircut}
    We calculate the pair fraction for subsamples of low mass and high mass major pairs by selecting only pairs that have separations less than some factor of the virial radius of its FoF group. 
    Note that a minimum separation limit of $10\kpc$ is always applied to all of our pairs (see Sec.~\ref{sec:methods-pairs}). 
    The virial radius is taken from the \texttt{Group\_R\_TopHat200} field in the group catalogs. 
   
    Major pairs typically correspond to the two most massive subhalos in their FoF group. 
    The virial mass of their FoF group is reasonably approximated by the combined subhalo mass of both galaxies in the pair.
    The combined mass of major low mass galaxy pairs recovers on average 98\% of the FoF group mass, and the combined mass of major high mass galaxy pairs recovers an average of 93\%. 
    For example, the mass of the Local Group is dominated by that of the MW and M31 \citep[e.g., ][]{Chamberlain2023}. 
    The virial radius of the FoF group is thus reasonably approximated by the virial radius of a halo with a virial mass equal to the combined subhalo mass of the two major pair members. 
    
    By focusing our separation criteria on the FoF group virial radius, the separation cut will vary both as a function of redshift and as a function of the combined mass of the pair consistently.
    For the high mass FoF groups, the median virial radius at $z=[0,1,2,3,4]$ is $\sim(348, 206, 134, 97, 76)\kpc$. 
    For the low mass FoF groups, the median virial radius at $z=[0,1,2,3,4]$ is $\sim[134, 85, 59, 43, 33]\kpc$.    
    
\begin{figure*}[htp]
        \centering
        \includegraphics[width=\textwidth]{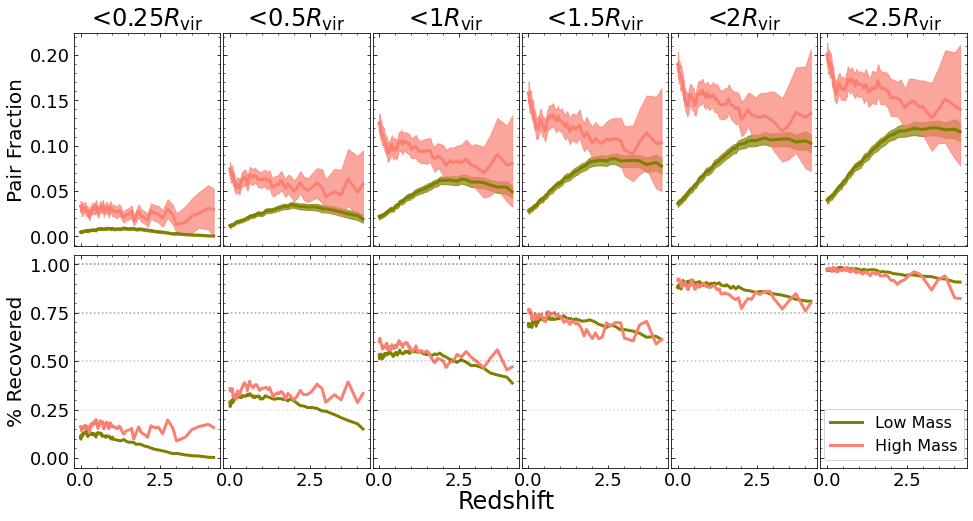}
        \caption{\label{fig:vircut}The median and 1st-99th percentile spread are shown by the solid lines and shaded regions, respectively. 
        (Top) The median pair fraction of the subset of high mass (pink) and low mass (green) pairs with 3D separations within a given factor of the pair's FoF group virial radius.
        Such a separation criteria will vary as a function of both the mass of the pair and the redshift. 
        Recovering the redshift evolution seen for the total sample (left panel of Figure~\ref{fig:pairfrac}) requires separation cuts of $r_{\mathrm{sep}} < 1.5\Rvir$ for low mass pairs and $r_{\mathrm{sep}} < 0.5\Rvir$ for high mass pairs. 
        Recovering the differences between high and low mass pair fraction trends seen for the total sample at $z>2$, requires $r_{\mathrm{sep}} < 0.5\Rvir$ for both galaxy types. 
        (Bottom) The recovery fraction, calculated as the fraction of the total collection of pairs recovered by the subset of pairs at the given separation cut. 
        We find that recovering the redshift evolution of the pair fractions of all galaxy pairs requires a separation cut such that the number of close pairs constitutes more than $\sim50\%$ of the full sample at all redshifts.
    }
    \end{figure*}

    We choose 6 subsamples, consisting of pairs with separations less than $0.25, 0.5, 1.0, 1.5, 2.0,$ and $2.5\Rvir$. 
    The top row of Fig.~\ref{fig:vircut} shows the median pair fraction (solid lines) for high mass (pink) and low mass (green) pairs in each of the 6 subsamples. 
    In the bottom row, we show the recovery fraction, which is the fraction of the total pair sample that is recovered by each subsample, i.e. the number of pairs that pass each separation criterion divided by the total number of pairs in the full sample presented in Sec.~\ref{sec:results-frac}.

    As the pair separation limits increase (left to right), the recovery fraction increases. So too does the pair fraction, as expected since each consecutive selection cut is less restrictive and will contain more of the full sample. 
    More than 75\% of the full pair sample is recovered at all redshifts when the sample contains all pairs within $2\Rvir$ (the two rightmost panels of Fig.~\ref{fig:vircut}).

    All subsamples of high mass pairs display broadly the same pair fraction redshift evolution as the full high mass sample (pink line in left panel of Figure~\ref{fig:pairfrac}): Each of the subsampled high mass pair fractions peak at $z=0$ and remain roughly constant or decrease at higher redshift. 
    The finer detail trends from the full sample, particularly the upturn at low $z$, are reasonably captured if $r_{\mathrm{sep}} < 0.5 \Rvir$, despite the recovery being $\sim$40\%. 

    The subsamples of low mass pairs, however, show a different trend from the full sample (green line in left panel of Figure~\ref{fig:pairfrac}) if the separation cut is too small. 
    For $r_{\mathrm{sep}} < 0.25\Rvir$, the pair fraction is flat or decreasing as a function of redshift (especially for $z>1$), similar to the behavior for high mass pairs. 
    For $r_{\mathrm{sep}} < 0.5\Rvir$ and $r_{\mathrm{sep}} < 1.0\Rvir$, the pair fraction rises to a peak at $z\sim2$ and then decreases. 
    When the separation is limited to $r_{\mathrm{sep}} < 1.5\Rvir$, the trends with redshift for the full low mass sample are recovered, particularly the roughly flat behavior from $z\sim2-4$, despite a sample recovery of $<75\%$. 
    We also find that differences between the redshift evolution of high mass pairs vs. low mass pairs (particularly at $z<2$) become apparent if $r_{\mathrm{sep}} < 0.5\Rvir$ for both galaxy types. 
    Overall, we find that recovering the redshift evolution of the pair fractions of all galaxy pairs requires a separation cut that contains at least the closest $\sim40-50\%$ of pairs from the full sample at all redshifts. 

\subsubsection{Pair separation limits based on static 3D physical separation} \label{sec:results-frac-sepcut}
    We calculate the pair fraction for subsamples of major low mass and high mass pairs by including only pairs with separations less than $[50,70,100,150,200,\mathrm{and\, }300]\kpc$.  
    Note that these separation criteria do not vary as a function of redshift or mass of the pair, and that a minimum separation limit of $10\kpc$ is always applied (see Sec.~\ref{sec:methods-pairs}).
    These separation criteria create subsamples of the \paircat\ containing low and high mass pairs with separations between $10-50\kpc$, $10-70\kpc$, and so on. 

\begin{figure*}[htp]
        \centering
        \includegraphics[width=\textwidth]{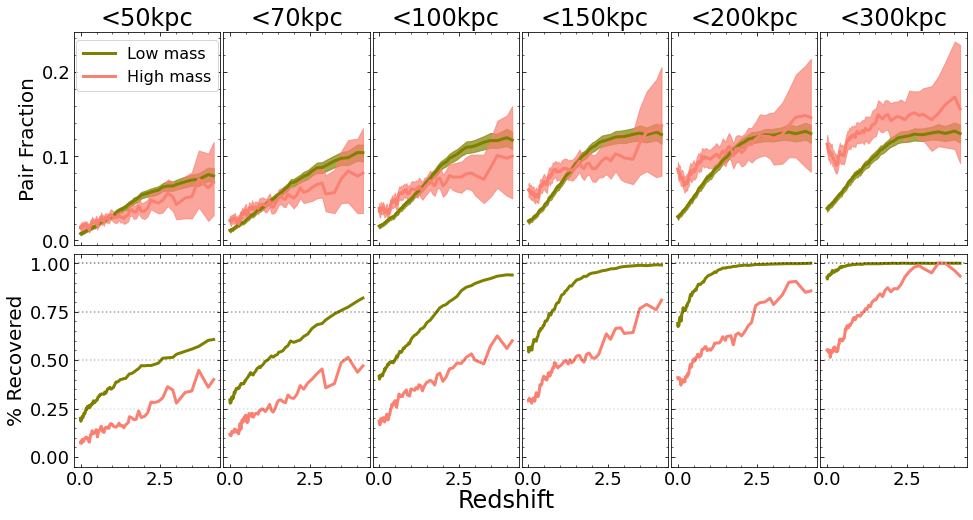}
        \caption{
        (Top) The major pair fraction of the subset of high mass (pink) and low mass (green) pairs with 3D separations less than the value given at the top of each column. 
        The median and 1st-99th percentile spread are shown by the solid lines and shaded regions, respectively. 
        The 3D separation cut does not vary as a function of redshift or mass of the system. 
        The general redshift evolution of the full high mass pair sample (decreasing with redshift for $z>1$) is not recovered for any physical separation cut, while the behavior of low mass pair fractions is well recovered by the $r_{\mathrm{sep}}<150\kpc$ subsample.
        The pair fractions for low mass and high mass pairs are virtually indistinguishable for separation cuts less than $70\kpc$. 
        (Bottom) The fraction of the total collection of pairs in the subset that passes each separation cut. 
        The recovered fraction varies as a function of redshift because a constant separation is chosen. 
        A higher fraction of low mass pairs is recovered than high mass pairs at all separation cuts. 
        Less than 50\% of all high mass pairs have separations $r_{\mathrm{sep}}<70\kpc$ at all redshifts, and $>50\%$ of the full high mass sample is recovered at all redshifts only for $r_{\mathrm{sep}}<300\kpc$. 
        Between 20-60\% of low mass pairs have separations $r_{\mathrm{sep}}<50\kpc$ at all redshifts, while a separation cut of $r_{\mathrm{sep}}<150\kpc$ captures roughly 50-100\% of the low mass pair population at all redshifts. 
        }
      \label{fig:sepcut}
    \end{figure*}

    Fig.~\ref{fig:sepcut} shows the median pair fractions for the subsamples of pairs with separations lower than the physical separation listed at the top of each column. 
    The bottom row of the plot shows the recovery fraction, or the number of pairs in each subsample compared to the full sample of major pairs. 
    Again, as the maximum separation increases (left to right), so too does the pair fraction and recovered fraction of the subsample. 
    
    The low mass major pair fraction (green) maintains roughly the same behavior (decreasing with decreasing redshift) for each separation cut, and first recovers the redshift evolution and magnitude of the pair fraction of the full sample (left panel of Fig.~\ref{fig:pairfrac}) for a separation cut of $r_{\mathrm{sep}} < 150\kpc$. 
    This is not surprising based on our results from the previous section as this separation is larger than the median virial radius of the sample and the recovery fraction is larger than 0.50 at all redshifts (reaching nearly 100\% at $z\sim3$).  
    The low mass pair fraction does not change significantly by excluding pairs with separations $>150\kpc$.

    On the other hand, none of the subsamples of the high mass pairs accurately recover the behavior of the full high mass sample. 
    In particular, the redshift evolution of the pair fraction between $z=0.25$ and 2.5 is not readily distinguishable from that of low mass pairs for any subsample from a physical separation cut.
    At higher redshifts, the median pair fraction increases rather than leveling off or decreasing as in the full sample.
    Even for $r_{\mathrm{sep}}<300\kpc$, the high mass pair fraction peaks at $z=4$, at odds with results using the full pair sample or any of the virial radius cuts in the previous section. 

    By applying a separation cut that is constant as a function of the redshift and mass of the system, the recovery fraction of the total sample varies markedly as a function of redshift.  
    This is in stark contrast to results from the previous section where the recovery fraction was roughly constant as a function of redshift.
    Because the same separation cut is applied to low and high mass pairs, the recovery fraction for low mass pairs is universally higher than for high mass pairs at each separation cut, which erases any differences in redshift evolution between the two galaxy types.
    
    
    \subsection{Comparison between TNG100-1 and TNG100-Dark}\label{sec:disc-dark}
    We repeated the entirety of our analysis using data from the TNG100-Dark simulation.
    Each step in our selection criteria was repeated identically, including our selection of group halos, abundance matching process, and selection of primary and secondary halos.
    We created an equivalent catalog to the \paircat\ containing only subhalo properties from the TNG100-Dark simulation, and calculated the low and high mass pair fractions for $z=0-4.2$. 

    We found that the redshift evolution of low mass and high mass pair fractions in the TNG100-Dark simulation mimic those of the TNG100 simulation. 
    In particular, low mass pair fractions remain flat between $z\sim2$ and 4, and decline from $z=2$ to a minimum at $z=0$. 
    High mass pair fractions likewise decrease with higher redshift for $z>1$, and peak at $z=0$.
    The numerical value of the pair fraction at all redshifts is $\sim10\%$ larger in TNG100-Dark for both low and high mass pairs and for major and minor pairs, which is likely a result of the suppressed abundance of low mass halos due to baryonic physics~\citep{chua17}. 


\section{Discussion}\label{sec:discussion}
We performed a pair fraction analysis for low mass and high mass pairs in the TNG100 simulation from $z=0-4.2$, utilizing the full spatial information that simulations enable to ensure that the pairs are physically colocated as part of the same FoF Group. 
In this section, we discuss the broader impacts and implications of the behavior of high and low mass pair fractions over time.

In Sec.~\ref{sec:disc-LG}, we draw comparisons between our pair sample and various Local Group pairs to quantify the likelihood of finding isolated analogs of such pairs at higher $z$.
In Sec.~\ref{sec:disc-comp}, we compare our results to previous studies of pair fractions that utilize simulations. 
We discuss the observational implications of the pair fraction behavior of low mass pairs and high mass pairs as a function of redshift
in Sec.~\ref{sec:disc-obs}.
We conclude by discussing some possibilities for the pair fraction difference between low and high mass pairs in Sec.~\ref{sec:disc-diff}.

   \subsection{Implications for Local Group Galaxies}\label{sec:disc-LG}
    Our study encompasses the stellar mass range and stellar mass ratios of a few well-studied pairs within the Local Group and Local Volume.
    Thus, we can make predictions for the frequency of finding isolated mass analogs of these systems at $z>0$ for the first time. 
    For example, isolated mass analogs of the following pairs are included in each specified sample:
    the LMC--SMC and NGC4490-4485\footnote{NGC4490-4485 is a low mass ($M_* = 7.2 \mbox{ and } 0.82 \times 10^9\Msun$, respectively) galaxy binary that is $\sim7\Mpc$ away from the MW~\citep{Theureau2007,Pearson2018}.} (low mass minor pair), the MW--LMC and M31--M33 (high mass minor pair), and the MW--M31 (high mass major pair).
    Note that both the NGC 4490--4485 and MW--M31 pairs pass all of our pair selection criteria, while the LMC--SMC,  MW--LMC, and M31--M33 pairs are non-isolated pairs within a more massive group environment.

    Mass analogs of the MW--M31 pair are most common today (pink solid line in the left panel of Fig.~\ref{fig:pairfrac}). 
    Given their large separation of $\sim$760 kpc, they would only be included in the $<2.5\Rvir$ panel of Figure~\ref{fig:vircut}. 

    We find that isolated mass analogs of the LMC--SMC and NGC 4490--4485 are roughly three times as common at $z>3$ than at $z=0$ (green dotted line in the right panel of Fig.~\ref{fig:pairfrac}). 
    These high $z$ pairs may enter a more massive group halo to become satellites of a larger galaxy~\citep[like the LMC--SMC; ][]{Besla2007,Patel2017a-Orbits}, or, if they remain isolated, may continue to merge in a manner similar to NGC 4490--4485~\citep{Pearson2018}. 
    
    Isolated mass analogs of the MW--LMC and M31--M33 systems are equally likely at $z>1$ as they are at present (right panel of Figure~\ref{fig:pairfrac}). 
    If we fold in the present-day separation of the MW--LMC system ($\sim50 \kpc$), this high mass minor pair becomes rare at low~$z$.\footnote{Fig.~\ref{fig:sepcut} displays pair fraction trends for major pairs. The equivalent plot for minor pairs, which we did not include here, shows the same trends as the major pairs, particularly in the $<50\kpc$ panel where the pair fraction for high mass minor pairs is $<0.02$ at $z=0$.}
    Specifically, we find that 2\% of MW's would host such a close LMC at $z=0$, which is consistent with other cosmological studies, such as \citet{Patel2017a-Orbits}, which finds that 3.8\% of MW-mass halos host an LMC-mass analog within 50 kpc at $z=0$. 
    We also find that such systems (high mass minor pairs with a low separation) are 2-3 times more common at high~$z$. M31--M33 analogs, which have higher separations $(\sim200\kpc)$, are more common than MW--LMC analogs at all redshifts.  

\subsection{Comparison to existing pair fraction studies}\label{sec:disc-comp}
    A direct comparison to pair fractions reported from observations is not straightforward, as observationally selected pairs suffer from contamination due to projected pairs and restrictive separation criteria that exclude more widely separated pairs.
    Instead, we compare our results to pair fractions and merger rates reported by other studies of the Illustris cosmological simulations to establish the reliability of our findings. 

        \citet{Besla2018} quantified low mass galaxy ($2\times10^{8} < M_{*} < 5\times10^{9}\,\Msun$) pair fractions at $z\sim0$ in the Illustris-1 hydrodynamic cosmological simulation, utilizing both projected and 3D pairs.
        The selection criteria for projected pairs include a projected separation cut, $r_{p} < 150\,\kpc$, and a relative line-of-sight velocity difference $\Delta v_{\rm los} < 150\,\kms$.
        
        Searching for pairs in a projected space while having access to the true 3D positions and velocities of the galaxies allowed \cite{Besla2018} to quantify the contamination fraction of false pairs due to projection effects. 
        They found that up to $\sim40$\% of identified companions were unrelated but appeared to be close due to projection effects. 
        The projected pair sample also enabled a direct comparison of the cosmologically-derived pair fractions with an equivalently selected low mass galaxy pair fraction from the Sloan Digital Sky Survey (SDSS).
        They found that the low mass major pair fraction from the SDSS sample was in good agreement with the simulations. The~\citet{Besla2018} study points to the Illustris simulation's ability to robustly constrain pair fractions of low mass galaxies, specifically at low $z$.

        We find reasonable agreement with the sample of physical 3D pairs in \cite{Besla2018}, which found a major pair fraction between 0.003-0.018 over a mass range of $2\times 10^8 < M_* < 5 \times 10^9$ $M_\odot$.
        Our roughly equivalent low mass major pair sample, when adopting a separation cut of $r_{\mathrm{sep}}<150\kpc$, results in a major pair fraction of $0.022^{+0.002}_{-0.003}$ at $z=0$. We believe our pair fractions are somewhat higher than in \cite{Besla2018} as our stellar mass range for primaries extends to a lower value of $10^8\Msun$. As such, our results for low mass pairs are robust.  

        For high mass pairs, \citet{Snyder2017} created mock catalogs using light cones in the Illustris-1 simulation to select pairs in the same fashion as done in observations. 
        They consider major pairs (stellar mass ratio $>1/4$) with: $1\times 10^{11}>\ms{1} >1\times 10^{10.5}\Msun$, projected distances between $14-71\kpc$, and a redshift separation of $\Delta z<0.02(1+z_{pri})$, corresponding to a velocity separation of $<1.8\times10^4\,\kms$ at $z=2$. 
        They find that the major pair fraction is constant or decreasing for $z>1$, which is in good agreement with observational studies.
        
        \citet{Snyder2023} extended this work, utilizing the TNG simulation to create mock images of extragalactic survey fields mimicking future planned surveys like JADES.
        From the mock images, they calculated the pair fraction for major pairs with projected separations between $5-70 \kpc$ and with redshift separations of $\Delta z< 0.02(1+z)$. 
        Again, they found a flat or decreasing major pair fraction with increasing redshift above $z=1$. 
        
        We cannot directly compare the values of the pair fractions in this work and \citet{Snyder2023}, since their work is done in projected space, but we can compare the trends as a function of redshift. 
        We find that the high mass major pair sample has a pair fraction that decreases with redshift above $z=1$, in agreement with \cite{Snyder2023}.

        In addition, \cite{RG2015} examine the merger rates of galaxies in the Illustris simulations as a function of stellar mass and redshift.
        From their Figures~7 and 10, there is roughly a factor of 4 difference in the major merger rate for high mass ($M_* \sim 10^{11} \Msun$) vs. low mass ($M_* \sim 10^8 \Msun$) galaxies at $z=0$, and the difference becomes smaller at higher redshift. 
        A strong mass dependence of the galaxy merger rate at $z=0$ is also seen in Figure~12 of \cite{Guzman-Ortega2023}, as well as in results from semi-empirical models \citep{Stewart2009, Hopkins2010}.
        From our Figure~\ref{fig:pairfrac}, there is also a factor of 4 difference in the pair fraction for high and low mass galaxies at $z=0$, which becomes smaller at higher redshift.
        The consistency between our results and the variation with stellar mass and redshift of the galaxy merger rate suggests that our finding of different pair fractions between low and high mass galaxies at $z=0$, as well as their different redshift evolution trends, is indeed reliable.  

\subsection{Implications for observational pair fraction studies}\label{sec:disc-obs}

    In Sec.~\ref{sec:results-frac}, we presented the pair fraction of high mass and low mass galaxy pairs and found that the relative frequency of the two populations evolves distinctly from $z=0-4.2$. 
    High mass pair fractions peak at $z=0$, decrease from $z=0$ to 0.3, after which they remain roughly constant or mildly decrease with increasing redshift. 
    In contrast, low mass pair fractions increase with increasing redshift until $z=2.5$, after which the frequency remains roughly constant with increasing redshift. 
    This behavior is seen for both major and minor pairs.  

    In Sec.~\ref{sec:results-frac-cuts}, we characterized the behavior of pair fractions for subsamples of the full pair catalog that pass additional separation cuts.
    We found that physical separation cuts which do not vary as a function of time and mass eliminate the ability to distinguish the different redshift evolution of high and low mass pair fractions. 
    Instead, by adopting a separation cut based on the virial radius of the group halo, we can accurately recover the different redshift evolution of low mass vs. high mass pair fractions, particularly at $z<2.5$.

    These results indicate that future observational studies that seek to compare low mass and high mass pair fractions, particularly as a function of time, must take care when determining their pair selection criteria.
    We advocate for separation criteria that varies as both a function of mass and redshift, such as our choice of $r_{\mathrm{sep}} < 1.0\Rvir$, where $\Rvir$ can be inferred using an estimate of the combined dark matter mass of the pair. 
    \footnote{Taking the observed stellar mass of each pair member, the halo mass for each pair can be estimated using the SMHM relation \citep[e.g.][]{Moster2013}. The combined halo mass of the pair can then be used as a proxy for the virial mass of the FoF group ($\rm M_{vir}$) to compute the virial radius, e.g. 
    \begin{equation}\label{eq}
    \Rvir = \sqrt[3]{\frac{3\rm M_{vir}}{4\pi \Delta_{c}\rho_{c}}},
    \end{equation} 
    where $\rho_c$ is the critical density of the Universe, and $\Delta_{c}$ is the overdensity constant~\citep[see][]{BinneyTremaine2008}. Both $\rho_c$ and $\Delta_{c}$ are functions of redshift, such that $\Rvir$ changes as a function of both the mass of the pair and redshift.}
    We have shown that utilizing fixed physical separation cuts can lead to significant deviation in the behavior of galaxy pair fractions for different galaxy masses and between $z=0$ and 4, and thus it is imperative to carefully consider the selection criteria used in future observational pair fraction studies of low and high mass pairs over time.
    
    In addition, our findings are specific to systems in isolated environments, and thus may not be representative of the pair fractions of a more "standard" observational field that contains pairs in isolated and high density environments.
    Mitigating this issue in observations would require making certain isolation cuts, such as those employed in \citet{Geha2013} to identify low mass pairs in isolated ($>1.5\,\Mpc$ away from an $\rm L^{*}$ galaxy) and non-isolated (within $1.5\,\Mpc$ of an $\rm L^{*}$ galaxy) environments.

    Given that chance projections will artificially boost the pair fraction as the pair separation is increased, recovering the general evolution of the pair fraction as a function of redshift may be more realistic than recovering the magnitude of the pair fraction itself, since the former requires smaller separation limits. 
    The requirement of $r_{\mathrm{sep}} < 1.0\,\Rvir$ for low mass pairs translates to median separation limits that are less than $\sim 60(30)\kpc$ at $z \sim2(4)$, which are sufficiently small as to avoid major projection effects. 
    For high mass pairs, this translates to separations less than $\sim135(75)$ at $z\sim2(4)$.

    Pair fractions are typically translated to galaxy merger rates using an observability window~\citep{Lotz2011}.
    The good agreement between the difference in the low mass and high mass pair fractions and the mass-dependent galaxy merger rates from \cite{RG2015} suggests that applying separation cuts which can recover this mass dependency are critical to reliably translating pair fractions to galaxy merger rates. 
    In fact, JWST is expected to identify low mass galaxies ($M_* > 10^8\Msun$) out to at least $z=10$~\citep{Cowley2018,Williams2018,Behroozi2020}. 
    We argue that the adopted pair separation criteria, particularly for low mass pairs, will greatly affect the measured pair fractions and consequently the inferred merger rates. 
    
\subsection{Underlying physical behavior of galaxy pairs}\label{sec:disc-diff}
    Though we find a distinct difference in the redshift evolution of low mass and high mass pair fractions, our study does not explicitly identify the specific driver of these differences.
    In a \lcdm\ universe, the buildup of structure proceeds hierarchically, with smaller structures merging to form larger and larger structures at later times.
    This may explain the significant decrease in the pair fraction of isolated low mass galaxies from $z=2$ to $z=0$, since more and more low mass systems would be accreted into larger group structures and would thus be removed from our isolated pair sample.
    
    On the other hand, the inherent dynamics of the pairs themselves may dictate the redshift evolution of their respective pair fractions.
    For example, if low mass pairs tend to have very short merger timescales, the total number of low mass pairs would decrease much more rapidly than massive pairs with a very long timescale.
    
    Understanding in further detail the origin of the discrepancy in the redshift evolution of low mass vs. high mass pairs requires tracking the specific pairs across simulation snapshots to trace their orbits and study changes in the mass and environment of the pairs across cosmic time.  
    We leave this as the focus of future work.

\section{Summary and Conclusions}\label{sec:summary}
In this paper, we construct a sample of low mass ($\rm 10^8<M_*<5\times10^9\,\Msun$) and high mass ($\rm 5\times10^9<M_*<10^{11}\,\Msun$) pairs from the TNG100 simulation from $z=0$ to 4.2. Pairs are selected as belonging to the same FoF group such that they are isolated and physically associated, and not contaminated by projection effects.
Major and minor pairs are determined using the stellar mass ratio (1--1/4 and 1/4--1/10, respectively) between the primary and secondary halo, where stellar masses are assigned using an abundance matching prescription from~\cite{Moster2013} to generate 1000 separate realizations of selected pairs at each redshift in order to sample the error space of the SMHM relation.

From this pair sample, we calculate the pair fraction as a function of redshift for 4 different pair types: low mass major pairs, low mass minor pairs, high mass major pairs, and high mass minor pairs. 
Our goal is to quantify the evolution of low mass and high mass galaxy pairs across cosmic time and to identify any differences in the redshift evolution of pair fractions of low mass and high mass pairs.
We also aim to determine how separation criteria for pair selection can affect the pair fraction evolution of high and low mass pairs.
To this end, we additionally employ both static and time- and mass-evolving separation cuts to explore the resulting impact on the pair fractions.

Our main findings are as follows:
\begin{itemize}
    \item The pair fraction for low mass and high mass pairs does not proceed identically throughout cosmic time.
    In fact, the two mass scales have opposite behaviors at $z<2.5$ (see Figure~\ref{fig:pairfrac}).
    The pair fraction of low mass pairs increases from $z=0$ to $z\sim2.5$, while the pair fraction of high mass pairs peaks at $z=0$ and is constant or slightly decreasing at $z>1$.
    \item At z=0, we find a low mass major (minor) pair fraction of $0.041^{+0.003}_{-0.004}$ ($0.034\pm0.004$), and a high mass major (minor) pair fraction of $0.207^{+0.012}_{-0.014}$ ($0.152^{+0.016}_{-0.017}$). 
    These results are consistent with other simulation-based studies that find that the merger rate for high mass pairs is $\sim4\times$ higher than for low mass galaxies \citep[e.g.,][]{RG2015}. 
    \item Low mass minor pairs evolve similarly to low mass major pairs as a function of redshift.
    \item The differences in redshift evolution of pair fractions for high and low mass pairs is well recovered using a subsample of pairs with a maximum separation cut that varies with mass and redshift, specifically for $r_{\mathrm{sep}}<1\Rvir$, where $\Rvir$ is the virial radius of the FoF group.
    This separation cut corresponds to separations less than [134(348), 59(134), 33(76)]$\kpc$ at z=[0,2,4] for low mass (high mass) pairs.
    In particular, a maximum separation cut of $r_{\mathrm{sep}}<1\Rvir$ encompasses between $\sim40-60\%$ of the full population of pairs at all redshifts, and recovers the distinct differences in the redshift evolution of low mass vs. high mass pair fractions from $z=0$ to 2.5. 
    \item The identified differences in the redshift evolution of high and low mass pairs are erased when a static physical separation cut is employed, i.e. a separation that does not evolve with redshift or with mass (e.g. $r_{\mathrm{sep}}<50\kpc$). 
    This occurs because fixed-separation cuts capture roughly 10-50\% more pairs in the low mass pair sample than the high mass sample at all redshifts.  
    Consequently, selecting the same volume to study the pair fractions for low and high mass galaxies via a static separation cut will bias inferred pair fractions. 
    \item Isolated mass-analogs of the MW--M31 are most common at $z=0$, while analogs of the MW--LMC and M31--M33 are equally common at $z=0$ as at $z>1$. 
    However, MW--LMC-type systems with very low separations ($\lesssim50\kpc$) are 2-3 times more common at higher redshift ($z\sim$2). 
    Isolated analogs of the LMC--SMC and  NGC 4490--4485 are 3 times more common for $z>3$. 
\end{itemize}
    
    A number of studies have identified a mass and redshift dependence in the galaxy merger rate as a function of time, particularly in cosmological simulations and semi-empirical models~\cite[see, e.g.][]{Stewart2009,Hopkins2010,RG2015}. 
    Our results show that galaxy pair fractions for physically associated, isolated pairs likewise evolve as a function of mass and redshift. 
    However, we also find that the recovered pair fractions are sensitive to the separation criteria that is used to define pairs.
    The redshift evolution of pair fractions for physically associated low mass and high mass galaxy pairs can only be distinguished when the separation criteria are a function of both mass and redshift.

    Observational campaigns that seek to recover low mass and high mass pair fractions should consider using mass- and redshift-evolving separation criteria to select pairs.
    Many observational pair fraction studies use static definitions of the maximum projected separation to determine close pairs, then translate between pair fractions and merger rates via an observability timescale. 
    The observability timescale, however, is not typically calibrated as a function of both mass and redshift, meaning that identical measured pair fractions of high and low mass pairs will yield indistinguishable merger rates. 
    
    In order to observationally recover the cosmologically expected difference between the merger rates of high and low mass galaxies, it is crucial to ensure that differences in the redshift evolution of the pair fraction for these galaxy populations can be adequately captured. 
    This will be of particular importance in the era of JWST, where more low mass ($M_*\sim10^8 \Msun$) galaxies and galaxy pairs will be discovered at higher redshift (up to $z>10$), and for future wide-field galaxy surveys enabled by the Roman Space Telescope and Rubin Observatory.


\section*{Acknowledgements}
K.C. would like to thank Greg Snyder for fruitful early conversations during this work. 
K.C. and G.B. are supported by NSF CAREER award AST-1941096. 
E.P. acknowledges financial support provided by NASA through grant number HST-GO-16628. Support for this work was also provided by NASA through the NASA Hubble Fellowship grant \#HST-HF2-51540.001-A awarded by the Space Telescope Science Institute, which is operated by the Association of Universities for Research in Astronomy, Incorporated, under NASA contract NAS5-26555.
Support for S.P. was provided by NASA through the NASA Hubble Fellowship grant \#HST-HF2-51466.001-A awarded by the Space Telescope Science Institute, which is operated by the Association of Universities for Research in Astronomy, Incorporated, under NASA contract NAS5-26555.
The work of S.S. is supported by the National Science Foundation Astronomy \& Astrophysics Research grant No. 1911107.

The \textsc{IllustrisTNG} simulations were undertaken with compute time awarded by the Gauss Centre for Supercomputing (GCS) under GCS Large-Scale Projects GCS-ILLU and GCS-DWAR on the GCS share of the supercomputer Hazel Hen at the High Performance Computing Center Stuttgart (HLRS), as well as on the machines of the Max Planck Computing and Data Facility (MPCDF) in Garching, Germany.

The National Radio Astronomy Observatory is a facility of the National Science Foundation operated under cooperative agreement by Associated Universities, Inc.

This work is based upon High Performance Computing (HPC) resources supported by the University of Arizona TRIF, UITS, and Research, Innovation, and Impact (RII) and maintained by the UArizona Research Technologies department.

We respectfully acknowledge the University of Arizona is on the land and territories of Indigenous peoples. Today, Arizona is home to 22 federally recognized tribes, with Tucson being home to the O’odham and the Yaqui. Committed to diversity and inclusion, the University strives to build sustainable relationships with sovereign Native Nations and Indigenous communities through education offerings, partnerships, and community service.

\bibliography{refs}{}
\bibliographystyle{aasjournal}

\end{document}

%% file: math_definitions.tex
\newcommand{\Msun}{\ensuremath{\mathrm{M}_\odot}}

\newcommand{\kms}{\ensuremath{\mathrm{km}~\mathrm{s}^{-1}}}

\newcommand{\kpc}{\ensuremath{\mathrm{\,kpc}}}
\newcommand{\Mpc}{\ensuremath{\mathrm{Mpc}}}

\newcommand{\tng}{\textsl{IllustrisTNG}} 
\newcommand{\lcdm}{\ensuremath{\Lambda \rm CDM}} 
\newcommand{\sublink}{\texttt{SUBLINK}} 
\newcommand{\subfind}{\texttt{SUBFIND}} 
\newcommand{\Mpeak}{\ensuremath{M_{\mathrm{peak}}}}
\newcommand{\Mhalo}{\ensuremath{M_{\mathrm{h}}}}